\documentclass[12pt,a4paper]{article}

\usepackage{amsfonts}
\usepackage{amsmath}
\usepackage{amssymb}

\newcommand\ld{\lambda}
\newcommand\Si{\Sigma}
\newcommand\De{\Delta}
\newcommand{\map}{\rightarrow}               % rightarrow (for maps)
\newcommand{\A}{{\hat A}}
\newcommand{\hP}{{\hat P}}
\newcommand{\cS}{{\cal S}}
\newcommand{\Hi}{{\cal H}}
\newcommand\BH{\mathcal{B(H)}}
\newcommand\PH{\mathcal{P(H)}}
\newcommand\PV{\mathcal{P}(V)}
\newcommand\TO{\mathbb{T}}   %truth object
\newcommand\bra[1]{\langle #1|\,}
\newcommand\ket[1]{\,|#1\rangle}
\newcommand\eq[1]{(\ref{#1})}
\newcommand\Ain[1]{A\,\varepsilon\,#1}
\newcommand\das[1]{\delta(\hat{#1})}            % daseinisation of #1,
\newcommand\dastoo[2]{\delta^o(\hat{#2})_{#1}}  % outer daseinisation of #2 to context #1
\newcommand\dasB[1]{\breve{\delta}(#1)}
\newcommand\spec[1]{{\rm sp}(\hat A)}
\newcommand\ps[1]{\underline{#1}}        % underline argument #1 (for presheaves)
\newcommand{\Om}{\ps{\Omega}}            % underlined Omega (presheaf of sieves, subobject classifier in cat of presheaves)
\newcommand{\Sig}{\ps{\Sigma}}            % Spectral presheaf
\newcommand{\SR}{\ps{{\bbR}^\succeq}}         % presheaf of real-valued order-reversing functions on V
\newcommand\Subcl[1]{{\rm Sub}_{{\rm cl}}(#1)} % clopen subobjects
\newcommand\Set{{\bf Set}}                    % Set
\newcommand\SetC[1]{\Set^{{#1}^{\rm op}}}      % presheaf cat
\newcommand\V[1]{{\cal V}(\Hi_{#1})}           % V(H), the context cat
\newcommand\fu[1]{\overline{#1}}							% overlining for (covariant) functors. REDEFINES OLDER \fu COMMAND!!!
\newcommand\Sgm{\fu{\Si}}											% the spectral functor for a von Neumann algebra N
\newcommand\SetBHop{\SetC{\V{}}}
\newcommand\SetBH{\Set^{\V{}}}
\newcommand\Q{\hat Q}
\newcommand\bbC{\mathbb{C}}
\newcommand\bbR{\mathbb{R}}
\newcommand\mc[1]{\mathcal{#1}}
\newcommand\wpsi{\ps{\mathfrak{w}}^{\psi}}

\begin{document}
\title{Topos Quantum Logic and Mixed States}
\author{Andreas D\"oring\footnote{andreas.doering@comlab.ox.ac.uk}\\
Computing Laboratory, University of Oxford\\}
\date{April, 2010}
\maketitle
%\begin{frontmatter}
%\title{Topos Quantum Logic and Mixed States}
%\author{Andreas D\"{o}ring}
%\address{Oxford University Computing Laboratory, Wolfson Building, Parks Road, Oxford OX1 3QD\\andreas.doering@comlab.ox.ac.uk}
%\small \textsf{a.doering@imperial.ac.uk}

\begin{abstract}
The topos approach to the formulation of physical theories includes a new form of quantum logic. We present this topos quantum logic, including some new results, and compare it to standard quantum logic, all with an eye to conceptual issues. In particular, we show that topos quantum logic is distributive, multi-valued, contextual and intuitionistic. It incorporates superposition without being based on linear structures, has a built-in form of coarse-graining which automatically avoids interpretational problems usually associated with the conjunction of propositions about incompatible physical quantities, and provides a material implication that is lacking from standard quantum logic. Importantly, topos quantum logic comes with a clear geometrical underpinning. The representation of pure states and truth-value assignments are discussed. It is briefly shown how mixed states fit into this approach.
\end{abstract}

%\begin{keyword}
%Topos approach, quantum logic, superposition, implication
%\end{keyword}
%\end{frontmatter}

%%%%%%%%%%%%%%%%%%%%%%%%%%%%%%%%%%%%%%%%%%%%%%%%%%%%%%%%%%%%%%%%%%%%%%%%%%%%%%%%%%%%%%%%%%%%%%%%%%%%%%%%%%%%%%%%%%%%%%
\section{Introduction}
At a very basic level, physics is about propositions of the form ``the physical quantity $A$ (of some given system $S$) has a value in the set $\Delta$ of real numbers'', written shortly as ``$A\varepsilon\Delta$''. One wants to know what truth-values such propositions have in a given state of the system. It is also of interest how the truth-value changes with the state (in time).

In classical physics, this is unproblematic: there is a space of states, and in any given (pure) state
\begin{itemize}
	\item[(a)] all physical quantities have a value,
	\item[(b)] all propositions of the form ``$A\varepsilon\Delta$'' have a truth-value.
\end{itemize}
For a classical system, propositions are represented by subsets of the state space. Usually, one restricts attention to Borel subsets, and we will follow this convention here. A (pure) state of a classical system is a point of the state space of the system,\footnote{We avoid the notion `phase space', which seems to be a historical misnomer.} and the truth-values of propositions are \textit{true}, \textit{false}. The Borel subsets of the state space form a Boolean $\sigma$-algebra, and the logic of classical physical systems is Boolean logic. Classical physics is a \emph{realist} theory fulfilling properties (a) and (b).

As is well-known, there is no such realist formulation of quantum theory: the Kochen-Specker theorem \cite{KS67,Doe05} shows that there is no state space of a quantum system analogous to the classical state space. Hilbert space does \emph{not} play this role. In particular, Kochen and Specker required that the physical quantities are represented as real-valued functions on the (hypothetical) state space of a quantum system and then showed that such a space does not exist. In fact, they showed the even stronger result that under very natural conditions it is impossible to assign values to all physical quantities at once, and hence it is also impossible to assign \textit{true} resp. \textit{false} to all propositions.

In standard quantum logic, which goes back to the seminal paper \cite{BvN36} by Birkhoff and von Neumann, propositions like ``$\Ain\De$'' are represented by projection operators on Hilbert space (via the spectral theorem). The projections form a non-distributive lattice, which makes the interpretation of the lattice operations as logical operations very dubious. Quantum logic is lacking a proper semantics. Pure states are represented by unit vectors in Hilbert space. Let $\hat E[A\varepsilon\Delta]$ denote the projection representing the proposition ``$\Ain{\Delta}$''. In a given state $\ket\psi$, we can calculate the probability of ``$A\varepsilon\Delta$'' being true in the state $\ket\psi$:
\begin{equation}
			P(\Ain{\Delta};\ket\psi):=\bra\psi\hat E[\Ain{\Delta}]\ket\psi\in[0,1].
\end{equation}
The interpretation is \emph{instrumentalist}: upon measurement of the physical quantity $A$, we will find the result to lie in $\Delta$ with probability $P(\Ain{\Delta};\ket\psi)$.

Standard quantum logic and its many generalisations \cite{DCG02} have a number of conceptual and interpretational problems. To us, non-distribuitivity and the fundamental dependence on instrumentalist notions seem the most severe. In the following, a proposal for a new form of quantum logic is sketched which overcomes these problems. This new form of quantum logic arose from the topos approach to the formulation of physical theories, initiated more than a decade ago by Butterfield and Isham \cite{IB98,IB99,IB00,IB02}. Further references are given below.

In Section 2, the basic structures of the topos approach are introduced. In Section 3, the representation of propositions and daseinisation of projections are discussed. Section 4 is concerned with pure states and how they serve to assign topos-internal truth-values to all propositions. Sections 3 and 4 are developing joint work with Chris Isham. A number of new small results is proved. In particular, it is shown that topos quantum logic `preserves' superposition. There is some emphasis on the conceptual discussion of the topos scheme. Section 5 sketches how mixed states can be treated in the topos approach and how this relates to the logical aspects. In Section 6, some related work is pointed out. Section 7 concludes.

%%%%%%%%%%%%%%%%%%%%%%%%%%%%%%%%%%%%%%%%%%%%%%%%%%%%%%%%%%%%%%%%%%%%%%%%%%%%%%%%%%%%%%%
\section{The basic structures}
In this section, some of the basic structures of the topos approach to quantum theory are introduced. We can only give a sketch and some intuitive ideas here, for a comprehensive presentation including many further aspects and results see \cite{DI(1),DI(2),DI(3),DI(4)} and \cite{DI08}. A short introduction to the topos approach in general is given in \cite{Doe07b}, and more recently in \cite{Doe10,Ish10}.

We assume that a quantum system is described by its algebra of observables. For simplicity, we assume that this algebra is $\BH$, the algebra of all bounded operators on a separable Hilbert space $\Hi$ of dimension $2$ or greater. The Hilbert space can be infinite-dimensional. To each physical quantity $A$ of the quantum system, there corresponds a self-adjoint operator $\A$ in $\BH$ and vice versa. $\BH$ is a \emph{von Neumann algebra}.

We emphasise that all our results generalise without extra effort to arbitrary von Neumann algebras.\footnote{It is no problem that physical quantities like position and momentum are described by unbounded operators, while a von Neumann algebra contains only bounded operators. An unbounded operator can be affiliated to a von Neumann algebra provided all its spectral projections lie in the algebra (see e.g. \cite{KR83a}).} 

%%%%%%%%%%%%%%%%%%%%%%%%%%%%%%%%%%%%%%%%%%%%%%%%%%%%%%%%%%%%%%%%%%%%%%%%%%%%%%%%%%%%%%
\subsection{The context category}
One central idea in the topos approach is to take \emph{contextuality} into account, as suggested by the Kochen-Specker theorem. For us, a \emph{context} is an abelian subalgebra of the non-abelian von Neumann algebra $\BH$. We consider only abelian von Neumann subalgebras, since we want enough projections in each algebra for the spectral theorem to hold. Moreover, we consider only those abelian subalgebras that contain the identity operator $\hat 1$ on $\Hi$. Let $\V{}$ denote the set of unital, abelian von Neumann subalgebras of $\BH$. $\V{}$ is a partially ordered set under the inclusion of smaller into larger algebras. If $V',V\in\V{}$ are contexts such that $V'$ is contained in $V$, we denote the inclusion as $i_{V'V}:V'\rightarrow V$. Since every poset is a category, we call $\V{}$ the \emph{context category}. Its objects are the abelian von Neumann subalgebras of $\BH$, and its arrows are the inclusions between them.

Every context $V\in\V{}$ provides a classical perspective on the quantum system, since, as in classical physics, all the physical quantities resp. the corresponding self-adjoint operators in a context commute. Of course, the perspective provided by a single context $V$ is partial, since the quantum system has non-commuting physical quantities that cannot all be contained in the context. But by taking \emph{all} contexts into account at once, and by moreover keeping track of their relations, one can hope to gain a complete picture of the quantum system.

%%%%%%%%%%%%%%%%%%%%%%%%%%%%%%%%%%%%%%%%%%%%%%%%%%%%%%%%%%%%%%%%%%%%%%%%%%%%%%%%%%%%%%
\subsection{The topos associated to a quantum system and its internal logic}			\label{Sec2.2}
Of course, it will not be enough to consider the context category $\V{}$ alone. Instead, one defines structures over $\V{}$ and relations between these structures. Concretely, one considers $\Set$-valued functors on the context category $\V{}$ and natural transformations between them.\footnote{The standard reference on category theory is \cite{McL71}.} At this point, a choice between covariant and contravariant functors must be made.

Let $V,V'\in\V{}$ such that $V'\subset V$. The step from the algebra $V$ to the smaller algebra $V'$ is a process of \emph{coarse-graining}: since $V'$ contains less self-adjoint operators and less projections than $V$, one can describe less physics from the perspective of $V'$ than from $V$. Mapping self-adjoint operators and projections from $V$ to $V'$ hence will make it necessary to approximate. If we consider a proposition ``$\Ain\De$'' about a physical quantity $A$ that is represented by a self-adjoint operator $\A$ in $V$, then there exists a projection $\hP=\hat E[\Ain\De]$ in $V$ that represents the proposition. It may happen that the smaller abelian subalgebra $V'$ does not contain the projection $\hP$. This means that the proposition ``$\Ain\De$'' cannot be stated from the perspective of $V'$. Both the projection $\hP$ and the corresponding proposition ``$\Ain\De$'' must be adapted to $V'$ by making them coarser. This leads to the idea of \emph{daseinisation}, which is discussed in detail in the following section.

On the other hand, the step from $V'$ to $V$ is trivial, since every self-adjoint operator and every projection in $V'$ is of course also contained in $V$. Going in this direction, one can merely embed the smaller algebra $V'$ into the larger one, without making use of the extra structure (more self-adjoint operators, more projections, hence more propositions) available in $V$.

This strongly suggests to use \emph{contra}variant functors on the context category $\V{}$: the arrows in $\V{}$ are the inclusions of smaller contexts $V'$ into larger ones like $V$, so if we want to incorporate coarse-graining, our functors over $\V{}$ must invert the direction of the arrows. The idea hence is to consider $\SetBHop$, the collection of $\Set$-valued contravariant functors---traditionally called \emph{presheaves}---over the base category $\V{}$. With natural transformations as arrows between the presheaves, $\SetBHop$ becomes a category. This category has all the extra structure that makes it into a \emph{topos}.\footnote{I.e., it has finite limits and colimits, exponentials as well as a subobject classifier.} Topos theory is a highly developed branch of pure mathematics, and we refer to the literature for more information \cite{Gol84,MM92,Jst02}. For us, the important aspect is that a topos is a category whose objects `behave like sets'. Each presheaf $\ps{\mc{P}}\in\SetBHop$ can be seen as kind of a generalised set, and each natural transformation $\tau:\ps{\mc{P}_1}\rightarrow\ps{\mc{P}_2}$ between presheaves is the analogue of a function between sets. Presheaves can have extra structure so as to become a group, a topological space, a ring etc. internally in the topos $\SetBHop$ (and natural transformations may or may not preserve this extra structure).

The context category $\V{}$ is the \emph{base category} of the topos $\SetBHop$, and the contexts $V\in\V{}$ are also called \emph{stages}. Each presheaf $\ps{\mc{P}}\in\SetBHop$ can be seen as a collection $(\ps{\mc{P}}_V)_{V\in\V{}}$ of sets, one for each context, together with functions $\ps{\mc{P}}(i_{V'V}):\ps{\mc{P}}_V\rightarrow\ps{\mc{P}}_{V'}$ whenever $V'\subset V$. (If $V'=V$, then the function $\ps{\mc{P}}(i_{VV})$ is the identity on $\ps{\mc{P}}_V$.)

The subobject classifier $\Omega$ in a topos is the object that generalises the set $\{0,1\}$ of truth-values (where $0$ is identified with \textit{false} and $1$ with \textit{true}) in the topos $\Set$ of sets and functions. In the topos $\SetBHop$, the subobject classifier $\Om$ is the presheaf of \emph{sieves} on $\V{}$. To each $V\in\V{}$, the set $\Om_V$ of all sieves on $V$ is assigned. A sieve $\sigma$ on $V$ is a collection of subalgebras of $V$ that is downwards closed, i.e., if $V'\in\sigma$ and $V''\subset V'$, then $V''\in\sigma$. The \emph{maximal sieve} on $V$ is just the downset $\downarrow\!\!V$ of $V$ in $\V{}$. If $V'\subset V$, then the function $\Om(i_{V'V}):\Om_V\rightarrow\Om_{V'}$ sends a sieve $\sigma\in\Om_V$ to the sieve $\sigma\cap\downarrow\!\!V'\in\Om_{V'}$.

A \emph{truth-value} in the internal logic of the topos $\SetBHop$ is a \emph{global element} $\gamma=(\gamma_V)_{V\in\V{}}$ of the subobject classifier $\Om$, i.e., we have $\gamma_V\in\Om_V$ for all $V\in\V{}$ and $\gamma_V\cap\downarrow\!\!V'=\gamma_{V'}$ whenever $V'\subset V$.

The intuitive interpretation of such a truth-value $\gamma$ simply is that for each context $V\in\V{}$, we have a local truth-value \textit{true} or \textit{false}: if $V\in\gamma_{\tilde V}$ for some $\tilde V\supseteq V$, then at $V$, we have \textit{true}, else we have \textit{false}. The fact that $\gamma$ is a global element guarantees that this is independent of the choice of $\tilde V$. Moreover, the fact that we have sieves means that if at some $V\in\V{}$ we have \textit{true}, then we have \textit{true} at all $V'\subset V$.

Physically, we interpret the contexts $V\in\V{}$ as classical perspectives on the quantum system under consideration. A truth-value $\gamma$ hence is contextual: it provides information about truth or falsity from each perspective $V$. Clearly, there is a truth-value $\gamma_1$ consisting of the maximal sieve $\downarrow\!\!V$ for each context. This is interpreted as \textit{totally true}, i.e., true from all perspectives $V\in\V{}$. The truth-value $\gamma_0$ consisting of the empty sieve for each $V$ is interpreted as \emph{totally false}. There are many other truth-values between $\gamma_0$ and $\gamma_1$. The truth-values are partially ordered under inclusion. It is well-known that they form a \emph{Heyting algebra}, the algebraic representative of \emph{intuitionistic} propositional logic. In particular, this means that conjunction and disjunction behave distributively. The main difference between an intuitionistic and a Boolean logical calculus is that the law of excluded middle need not hold in the former. If $H$ is a Heyting algebra with top element $1$, and $a\in H$ with $\neg a$ its negation, then
\begin{equation}
			a\vee\neg a\leq 1.
\end{equation}
In a Boolean algebra, equality holds.

The internal logic provided by the topos $\SetBHop$ hence is distributive, multi-valued, contextual and intuitionistic. We will apply this logical structure to quantum theory.

%%%%%%%%%%%%%%%%%%%%%%%%%%%%%%%%%%%%%%%%%%%%%%%%%%%%%%%%%%%%%%%%%%%%%%%%%%%%%%%%%%%%%%
\subsection{The spectral presheaf}
The fact that we are using Boolean logic in classical physics is closely tied to the fact that classical physics is based upon the idea of a \emph{state space} $\mc{S}$. A proposition ``$\Ain\De$'' about a physical quantity $A$ of the system is represented by a subset $S$ of the state space. This subset contains all states (i.e., elements of the state space) in which the proposition is true. Usually, one does not consider all subsets of state space, but restricts attention to measurable ones. The Borel subsets $B(\mc{S})$ form a $\sigma$-complete Boolean algebra.

In classical physics, in any given state $s\in\mc{S}$, every proposition has a truth-value. If $s$ lies in the subset of $\mc{S}$ representing the proposition, then the proposition is \textit{true}, otherwise it is \textit{false}.

The Kochen-Specker theorem \cite{KS67} shows that there is no analogous state space picture for quantum theory. The theorem is often interpreted as meaning that there are no non-contextual truth-value assignments in quantum theory. The topos approach takes this as a motivation and starting point. For each context $V\in\V{}$, there exists a state space picture similar to the classical case: each $V$ is an abelian $C^*$-algebra, so by Gel'fand duality there is an isomorphism $\mc{G}:V\rightarrow C(\Sigma_V)$ of $C^*$-algebras between $V$ and the continuous, complex-valued functions on the Gel'fand spectrum $\Sigma_V$ of $V$. Here, the Gel'fand spectrum $\Sigma_V$, which is a compact Hausdorff space, takes the r\^ole of the state space for the physical quantities described by self-adjoint operators in $V$. Each self-adjoint operator $\A\in V$ is sent to the real-valued function $\mc{G}(\A)$ on $\Sigma_V$, given by
\begin{equation}
			\forall \ld\in\Sigma_V:\mc{G}(\A)(\ld)=\ld(\A.)
\end{equation}
It holds that $\rm{im}(\mc{G}(\A))=\spec{\A}$. Since $V$ is a von Neumann algebra, the Gel'fand spectrum $\Sigma_V$ is extremely disconnected.

The main idea is to define a presheaf $\Sig$ over the context category $\V{}$ from all the local state spaces $\Sigma_V,\;V\in\V{},$ by assigning to each context $V\in\V{}$ its Gel'fand spectrum $\Sig_V=\Sigma_V$. If $V'\subset V$, then there is a canonical function
\begin{eqnarray}
			\Sig(i_{V'V}):\Sig_V &\longrightarrow& \Sig_{V'}\\			\nonumber
			\ld &\longmapsto& \ld|_{V'}.
\end{eqnarray}
This defines the \emph{spectral presheaf} $\Sig$, which is the analogue of the state space $\mc{S}$ of a classical system. It is a generalised set in the sense discussed in section \ref{Sec2.2}.

Each context $V\in\V{}$ is determined by its lattice of projections $\PV$ (since a von Neumann algebra is generated by its projections). The projections $\Q\in\PV$ represent propositions ``$\Ain\De$'' that can be made from the perspective of $V$. Since $V$ is an abelian von Neumann algebra, the projection lattice $\PV$ is a distributive lattice. Moreover, $\PV$ is complete and orthocomplemented. There is a lattice isomorphism between $\PV$ and $\mc{C}l(\Sig_V)$, the lattice of \emph{clopen}, i.e., closed and open subsets of the Gel'fand spectrum $\Sig_V$ of $V$:
\begin{eqnarray}			\label{alpha}
			\alpha:\PV &\longrightarrow& \mc{C}l(\Sig_V)\\			\nonumber
			\hP &\longmapsto& S_\hP:=\{\ld\in\Sig_V \mid \ld(\hP)=1\}.
\end{eqnarray}
Locally, at each $V\in\V{}$, this gives the correspondence between projections in $V$ and subsets of the Gel'fand spectrum $\Sig_V$.

%%%%%%%%%%%%%%%%%%%%%%%%%%%%%%%%%%%%%%%%%%%%%%%%%%%%%%%%%%%%%%%%%%%%%%%%%%%%%%%%%%%%%%%
\section{Representation of Propositions}
\subsection{Daseinisation of Projections}
Let $S$ be a given quantum system, and let ``$\Ain\De$'' be a proposition about the value of some physical quantity $A$ of the system. The task is to find a suitable representative of the proposition within the topos scheme.

The main idea is very simple: the spectral presheaf $\Sig$ is an analogue of the state space of a classical system. Since, in classical physics, propositions correspond to Borel subsets of the state space, we construct suitable subsets, or rather, subobjects, of the spectral presheaf that will serve as representatives of propositions.

There is a straightforward way of doing this: let ``$\Ain\De$'' be a proposition, and let $\hP=\hat E[\Ain\De]$ be the corresponding projection in $\PH$. In a first step, we `adapt' the projection $\hP$ to all contexts by defining
\begin{equation}
			\forall V\in\V{}:\dastoo{V}{P}:=\bigwedge\{\Q\in\PV\mid\Q\geq\hP\}.
\end{equation}
That is, we approximate $\hP$ from above by the smallest projection in $V$ that is larger than or equal to $\hP$. On the level of local propositions\footnote{We call those propositions \emph{local (at $V$)} that are represented by projection operators in $V$ via the spectral theorem. The proposition ``$\Ain\De$'' that we want to represent is \emph{global}. For each context $V\in\V{}$, the global proposition becomes coarse-grained to give some local proposition.}, we pick the \emph{strongest} local proposition implied by ``$\Ain\De$'' that is available from the perspective of $V$. In simple cases, $\dastoo{V}{P}$ represents a local proposition ``$\Ain\Gamma$'', where $\Gamma\supseteq\De$. In general, the self-adjoint operator $\A$ representing a physical quantity $A$ need not be contained in $V$ and thus the proposition represented by $\dastoo{V}{P}$ is of the form ``$B\in\Gamma$'', where $B$ is a physical quantity such that the corresponding self-adjoint operator $\hat B$ is in $V$.\footnote{We remark that even if $\A\notin V$, we can still have $\hP=\hat E[\Ain\De]\in V$.} In any case, $\dastoo{V}{P}\geq\hP$.

The central conceptual idea in the definition of the representative of a proposition ``$\Ain\De$'' is coarse-graining. Each context $V\in\V{}$ provides a classical perspective on the quantum system, characterised by the collection $\PV$ of projection operators in $V$. The projections in $V$ correspond to local propositions about the values of physical quantities in $V$. If the global proposition ``$\Ain\De$'' that we start from is a proposition about some physical quantity $A$ that is is represented by a self-adjoint operator $\A$ in $V$, then the projection $\hP$ representing ``$\Ain\De$'' is also contained in $V$, and daseinisation will pick this projection at $\hP$ (i.e., $\dastoo{V}{P}=\hP$). If $\A\notin V$ and $\hP\notin V$, we have to adapt $\hP$ to the context $V$. It is natural to pick the strongest local proposition implied by ``$\Ain\De$'' that can be made from the perspective of $V$. On the level of projections, this means that one has to take the smallest projection in $V$ larger than $\hP$. In this case, $\dastoo{V}{P}>\hP$.

From $\hP$, we thus obtain a collection of projections, one for each context $V\in\V{}$. We then use, for each $V\in\V{}$, the isomorphism \eq{alpha} to obtain a family $(S_{\dastoo{V}{P}})_{V\in\V{}}$ of clopen subsets. It is straightforward to show that for all $V',V\in\V{}$ such that $V'\subset V$, it holds that
\begin{equation}
			S_{\dastoo{V}{P}}|_{V'}=\{\ld|_{V'} \mid \ld\in S_{\dastoo{V}{P}}\}\subseteq S_{\dastoo{V'}{P}}
\end{equation}
(see Thm. 3.1 in \cite{DI(2)}. Actually, there it is shown that equality holds, which is more than we need here.) This means that the family $(S_{\dastoo{V}{P}})_{V\in\V{}}$ forms a subobject---which is nothing but a subpresheaf---of the spectral presheaf $\Sig$. This subobject will be denoted as $\ps{\das{P}}$ and is called the \emph{daseinisation of $\hP$}.

While we defined the subobject $\ps{\das{P}}$ of $\Sig$ stagewise, i.e., for each $V\in\V{}$, the subobject itself is a global object, consisting of all the subsets
\begin{equation}
			\ps{\das{P}}_V=S_{\dastoo{V}{P}}
\end{equation}
for $V\in\V{}$, and the functions
\begin{equation}
			\ps{\das{P}}_V\longrightarrow\ps{\das{P}}_{V'},\ \ \ \ \ \ld\longmapsto\ld|_{V'}
\end{equation}
between them (for all $V'\subseteq V$). In other words, $\ps{\das{P}}$ is a presheaf over the context category $\V{}$ and not a mere set. The whole of $\ps{\das{P}}$ represents the proposition ``$\Ain\De$'' (where $\hP$ is the projection corresponding to the proposition ``$\Ain\De$''). Many mathematical arguments concerning subobjects can be made stage by stage, yet the global character of subobjects is important both mathematically and in the physical interpretation.

A subobject $\ps S$ of $\Sig$ such that the components $\ps S_V$ are clopen sets for all $V$ is called a \emph{clopen subobject}. One can show that the clopen subobjects form a complete Heyting algebra $\Subcl{\Sig}$ (see Thm. 2.5 in \cite{DI(2)}). The subobjects obtained from daseinisation are all clopen. Compared to all subobjects of $\Sig$, the use of clopen ones has some technical advantages. We regard the Heyting algebra $\Subcl{\Sig}$ of clopen subobjects of the spectral presheaf as the algebra representing (propositional) quantum logic in the topos formulation. $\Subcl{\Sig}$ is the analogue of the Boolean $\sigma$-algebra of Borel subsets of the state space $\mc{S}$ of a classical system.

In the following, we discuss the main properties of daseinisation and of topos quantum logic in general.

%%%%%%%%%%%%%%%%%%%%%%%%%%%%%%%%%%%%%%%%%%%%%%%%%%%%%%%%%%%%%%%%%%%%%%%%%%%%%%%%%%%%%%%%%%%%%%%%%%
\subsection{Properties of daseinisation and their physical interpretation}
The following mapping is called \emph{daseinisation of projections}:
\begin{eqnarray}
		\ps\delta:\PH &\longrightarrow& \Subcl{\Sig}\\			\nonumber
		\hP &\longmapsto& \ps{\das{P}}.
\end{eqnarray}
It is straightforward to show that daseinisation has the following properties:
\begin{itemize}
	\item[(1)] If $\hP<\Q$, then $\ps{\delta(\hP)}<\ps{\delta(\Q)}$, i.e., daseinisation is order-preserving;
	\item[(2)] the mapping $\ps{\delta}:\PH\rightarrow\Subcl{\Sig}$ is injective, that is, two inequivalent propositions\footnote{It is well-known that the mapping from propositions to projections is many-to-one. Two propositions are equivalent if they correspond to the same projection.} correspond to two different subobjects;
	\item[(3)] $\ps{\delta(\hat 0)}=\ps 0$, the empty subobject, and $\ps{\delta(\hat 1)}=\Sig$. The trivially false proposition is represented by the empty subobject, the trivially true proposition is represented by the whole of $\Sig$.
\end{itemize}	
Moreover, we will show that
\begin{itemize}
	\item[(4)] for all $\hP,\Q\in\PH$, it holds that $\ps{\delta(\hP\vee\Q)}=\ps{\delta(\hP)}\vee\ps{\delta(\Q)}$, that is, daseinisation preserves the disjunction (Or) of propositions;
	\item[(5)] for all $\hP,\Q\in\PH$, it holds that $\ps{\delta(\hP\wedge\Q)}\leq\ps{\delta(\hP)}\wedge\ps{\delta(\Q)}$, that is, daseinisation does not preserve the conjunction (And) of propositions; 
	\item[(6)] in general, $\ps{\delta(\hP)}\wedge\ps{\delta(\Q)}$ is not of the form $\ps{\delta(\hat R)}$ for a projection $\hat R\in\PH$, and daseinisation is not surjective.
\end{itemize}
Daseinisation can be seen as a `translation' mapping between ordinary, Birk-hoff-von Neumann quantum logic \cite{BvN36}, which is based upon the non-distributive lattice of projections $\PH$ in $\BH$, and the topos form of propositional quantum logic, which is based upon the distributive lattice $\Subcl{\Sig}$. The latter more precisely is a Heyting algebra.

The properties (1--3) clearly are physically sensible. Before discussing properties (4--6) in some more detail, we emphasise that this representation of propositions by subobjects of the spectral presheaf, an object in the topos $\SetBHop$, has a strong geometric aspect. The spectral presheaf is the object that naturally incorporates the state spaces of all abelian subalgebras $V\in\V{}$ of the algebra $\BH$ of physical quantities of the quantum system. Moreover, the local state spaces $\Sig_V$ are related by the canonical restriction functions $\Sig(i_{V'V}):\Sig_V\rightarrow\Sig_{V'},\;\ld\mapsto\ld|_{V'},$ for all $V',V\in\V{}$ such that $V'\subset V$.

The spectral presheaf can be seen as a topological space in the topos $\SetBHop$, and it is closely related to the internal Gel'fand spectrum of an abelian $C^*$-algebra $\fu{\BH}$ in the functor topos $\SetBH$ that can be defined canonically from $\BH$ as suggested in \cite{HLS09}. Details about the relation between the spectral presheaf $\Sig$ and the spectrum of $\fu{\BH}$ can be found in \cite{Doe10b}. This strong geometrical and topological character of our quantum state space $\Sig$ is very different from the usual interpretation of Hilbert space as a state space, with closed subspaces resp. the projections onto these as representatives of propositions. In particular, neither the spectral presheaf $\Sig$ nor its components $\Sig_V,\;V\in\V{},$ are linear spaces.

In order to prove property (4), we first observe that for every $V\in\V{}$, the mapping
\begin{eqnarray}
			\delta^o_V:\PH &\longrightarrow& \PV\\			\nonumber
			\hP &\longmapsto& \bigwedge\{\Q\in\PV \mid \Q\geq\hP\}
\end{eqnarray}
is order-preserving. Let $\hP,\Q\in\PH$, then $\dastoo{V}{P}\leq\delta^o(\hP\vee\Q)_V$ and $\dastoo{V}{Q}\leq\delta^o(\hP\vee\Q)_V$, so $\dastoo{V}{P}\vee\dastoo{V}{Q}\leq\delta^o(\hP\vee\Q)_V$. Conversely, $\dastoo{V}{P}\vee\dastoo{V}{Q}\geq\hP$ and $\dastoo{V}{P}\vee\dastoo{V}{Q}\geq\Q$, so $\dastoo{V}{P}\vee\dastoo{V}{Q}\geq\hP\vee\Q$. But since $\dastoo{V}{P}\vee\dastoo{V}{Q}\in\PV$ and $\delta^o(\hP\vee\Q)_V$ is the smallest projection in $V$ larger than or equal to $\hP\vee\Q$, we also have $\dastoo{V}{P}\vee\dastoo{V}{Q}\geq\delta^o(\hP\vee\Q)_V$. Since the join of subobjects is defined stagewise, property (4) follows.

It is easy to see that property (4), the preservation of joins, can actually be generalised to arbitrary joins,
\begin{equation}
			\ps{\delta}(\bigvee_{i\in I}\hP_i)=\bigvee_{i\in I}\ps{\delta(\hP_i)}.
\end{equation}
The join of projections relates to \emph{superposition} in standard quantum logic. The following argument is from standard quantum logic: let ``$\Ain\De$'' and ``$B\varepsilon\Gamma$'' be two propositions, represented by projections $\hP$ resp. $\Q$. Any unit vector $\psi$ in the closed subspace $\hP\Hi$ of Hilbert space represents a pure state in which the proposition ``$\Ain\De$'' is true (i.e., the expectation value of $\hP$ in such a state is $1$). Similarly, every unit vector in $\Q\Hi$ is a state such that ``$B\varepsilon\Gamma$'' is true. 

\paragraph{Superposition without linearity.} The join $\hP\vee\Q$ of the two projections is the projection onto the closure of the linear subspace spanned by $\hP\Hi$ and $\Q\Hi$. In general, there are unit vectors, i.e., pure states $\psi$ in the closed subspace $(\hP\vee\Q)\Hi$ that are neither in $\hP\Hi$ nor in $\Q\Hi$. Such a state $\psi$ makes the proposition ```$\Ain\De$ or $B\varepsilon\Gamma$'', represented by $\hP\vee\Q$, true, despite the fact that in the state $\psi$, neither ``$\Ain\De$'' nor ``$B\varepsilon\Gamma$'' are true. A state $\psi$ of this kind can be written as a linear combination of vectors in $\hP\Hi$ and $\Q\Hi$ and is called a superposition state. The fact that states can be superposed by linear combinations is a fundamental fact of quantum theory. Clearly, superposition relates directly to the fact that Hilbert space is a linear space.

In this argument, one uses a certain feature of standard quantum logic that has been regarded as problematic: the fact that closed subspaces or projections represent physical properties in an \emph{intensional sense} and, at the same time, are the \emph{extensions} thereof, namely the collection of states which make the proposition true. This extensional collapse has been called the ``metaphysical disaster'' of standard quantum logic by Foulis and Randall \cite{RF83}, see also the discussion in \cite{DCG02}, where the problem is stated as:

\small ``The standard structures seem to determine a kind of extensional collapse. In fact, the closed subspaces of a Hilbert space represent at the same time physical properties in an intensional sense and the extensions thereof (sets of states that certainly verify the properties in question). As happens in classical set theoretical semantics, there is no mathematical representative for physical properties in an intensional sense. Foulis and Randall have called such an extensional collapse ``the metaphysical disaster'' of the standard quantum logical approach.''
\normalsize

In our topos approach, we did not invoke states yet, nor is our quantum state object $\Sig$ a linear space. Yet, we have the remarkable fact that daseinisation `translates' the disjunction of projections into the disjunction of clopen subobjects, i.e., daseinisation is a join-semilattice morphism. Interestingly, the binary join in $\Subcl{\Sig}$ is defined componentwise by set-theoretic union: let $\ps S_1,\ps S_2$ be two clopen subobjects, then
\begin{equation}
			\forall V\in\V{}:(\ps S_1\vee\ps S_2)_V=\ps S_{1;V}\cup\ps S_{2;V}.
\end{equation}
Differently from the Hilbert space situation, the linear span of linear subspaces does not play any role. The behaviour of projections under joins, which in standard quantum theory is so closely linked to superposition and the linear character of Hilbert space, is mapped by daseinisation to a lattice where joins are given by set-theoretic unions (in each component). Despite the fact that the spectral presheaf is not a linear space, daseinisation thus preserves a central aspect of standard quantum logic, namely that part which relates to superposition.

We further remark that we avoid the metaphysical disaster criticised by Foulis and Randall. Clopen subobjects represent propositions, but they are \emph{not} collections of states that make the propositions true.

\paragraph{Conjunction and coarse-graining.} Property (5) shows that conjunction of projections is not preserved. It is straightforward to construct a counterexample: let $\hP\in\PH$ be a projection, and let $V\in\V{}$ be a context that does not contain $\hP$. Then $\dastoo{V}{P}>\hP$ and $\delta^o(\hat 1-\hP)_V>\hat 1-\hP$, so $\dastoo{V}{P}\wedge\delta^o(\hat 1-\hP)_V>\hat 0$, while of course $\hP\wedge(\hat 1-\hP)=\hat 0$. Since the meet of subobjects is defined stagewise, property (5) follows. This clearly also implies property (6).

The counterexample shows that preservation of conjunction does \emph{not} fail just because of non-commutativity of the projections, which usually is interpreted as expressing incompatibility of the propositions represented by the projections. Rather, in the counterexample preservation of conjunction is not given due to coarse-graining. 

Dalla Chiara and Giuntini \cite{DCG02} sum up another common criticism of the standard quantum logic formalism:

\small ``The lattice structure of the closed subspaces automatically renders the quantum proposition system closed under logical conjunction. This seems to imply some counterintuitive consequences from the physical point of view. Suppose two experimental propositions that concern two strongly incompatible quantities, like ``the spin in the $x$ direction is up'', ``the spin in the $y$ direction is down''. In such a situation, the intuition of the quantum physicist seems to suggest the following semantic requirement: the conjunction of our propositions has no definite meaning; for, they cannot be experimentally tested at the same time. As a consequence, the lattice proposition structure seems to be too strong.''
\normalsize

In the topos approach, as in standard quantum logic, the conjunction of any two propositions is defined, but there is an interesting conceptual twist: the built-in contextuality and coarse-graining take care of the fact that there are strongly incompatible propositions as the ones mentioned above. Let $\hP$ be the projection representing the proposition ``the spin in the $x$ direction is up'' (which clearly is of the form ``$\Ain\De$''), and let $\Q$ represent ``the spin in the $y$ direction is down''. Then $\hP$ and $\Q$ do not commute, so they are not both contained in any context $V\in\V{}$. If we consider a context $V$ such that $\hP\in V$, then $\Q\notin V$ and hence $\dastoo{V}{Q}>\Q$. From a perspective of such a context $V$, the proposition ``the spin in the $y$ direction is down'' becomes coarse-grained, potentially to become the trivially true proposition represented by the identity operator $\hat 1$. Similarly, if a context $V$ contains $\Q$, then $\dastoo{V}{P}>\hP$, and hence the proposition ``the spin in the $x$ direction is up'' becomes coarse-grained from the perspective of such a context. There is no single context that allows to express the conjunction between the incompatible propositions within the context. But, since the (clopen) subobjects representing propositions are global objects, it still makes sense to talk about the conjunction of incompatible propositions in the topos scheme.

When discussing states and how they assign truth-values to propositions, we will see that there are non-trivial propositions that are not totally true in \emph{any} state. This is possible because the topos approach provides us with the collection of all clopen subobjects of the spectral presheaf as representatives of propositions. Among them are many that are not of the form $\ps{\das{P}}$ for a projection $\hP$. In contrast to that, in standard quantum logic every non-trivial proposition corresponds to a non-trivial closed subspace of Hilbert space, so there always are states that make the proposition true.

\paragraph{Material implication.} Each Heyting algebra $H$ has an implication, given by
\begin{equation}
			\forall x,y\in H : (x\Rightarrow y)=\bigvee\{z\in H \mid z\wedge x \leq y\}.
\end{equation}
Applied to our Heyting algebra $\Subcl{\Sig}$, whose elements represent propositions about the quantum system under consideration, this becomes
\begin{equation}
			\forall \ps S_1,\ps S_2\in\Subcl{\Sig} : (\ps S_1\Rightarrow\ps S_2)=\bigvee\{\ps S\in\Subcl{\Sig} \mid \ps S\wedge\ps S_1\leq\ps S_2\}.
\end{equation}
Using the well-known form for Heyting implication in presheaf topoi (see e.g. \cite{MM92}, p56), this can be evaluated concretely for all $V\in\V{}$ as
\begin{equation}
			(\ps S_1\Rightarrow\ps S_2)_V=\{\ld\in\Sig_V \mid \forall V'\subseteq V:\text{ if }\ld|_{V'}\in\ps S_{1;V'} \text{ then }\ld|_{V'}\in\ps S_{2;V'}\}.
\end{equation}
Note that this expression is not local at $V$, since a local definition would fail to give a subobject.

Hence, topos quantum logic comes with a material implication. This is another improvement compared to standard quantum logic, which is suffering from the lack of a proper implication. In particular, the Sasaki hook does not provide a material implication, as is well known.

\paragraph{Negation in topos quantum logic.} The negation is given in terms of the Heyting implication as usual:
\begin{equation}
			\neg\ps S:=(\ps S\Rightarrow \ps 0),
\end{equation}
where $\ps 0$ is the minimal element in the Heyting algebra $\Subcl{\Sig}$, namely the empty subobject.

This can be evaluated concretely for all $V\in\V{}$ as
\begin{equation}
			\neg\ps S_V:=\{\ld\in\Sig_V \mid \forall V'\subseteq V : \ld|_{V'}\notin \ps S_{V'}\}.
\end{equation}
%-- Distributivity

%-- other way of constructing subobjects: pullbacks

\section{Pure states and truth-value assignments}
\subsection{Truth objects and pseudo-states}
Let $\psi\in\Hi$ be a unit vector. As usual, $\psi$ is identified with the \emph{vector state} it determines:
\begin{eqnarray}
			w_\psi:\BH &\longrightarrow& \bbC\\			\nonumber
			\A &\longmapsto& w_\psi(\A)=\bra\psi\A\ket\psi.
\end{eqnarray}
The vector state $\psi$ is a pure state on $\BH$, i.e., an extreme point of the space of states (positive linear functionals of norm $1$) on $\BH$. In the topos approach, one cannot simply pick a (global) element of the spectral presheaf $\Sig$ as the representative of a state, since $\Sig$ has no global elements at all. As Butterfield and Isham observed, this is exactly equivalent to the Kochen-Specker theorem \cite{IB98,IB99,IB00,IB02}.

Instead, one defines a presheaf $\ps\TO^\psi$ over $\V{}$ that collects all those propositions that are totally true in the state $\psi$. For each $V\in\V{}$, let
\begin{eqnarray}			\label{DefTO}
			\ps\TO^\psi_V &=& \{S_\hP\in\mc{C}l(\Sig_V) \mid \hP\geq\hP_\psi\}.
\end{eqnarray}
Here, $S_\hP=\alpha(\hP)$ is the clopen subset of $\Sig_V$ corresponding to $\hP$ as defined in \eq{alpha}, and $\hP_\psi$ is the projection onto the one-dimensional subspace of Hilbert space determined by $\psi$. The component $\ps\TO^\psi_V$ hence contains all those clopen subsets of $\Sig_V$ that (a) represent local propositions that can be made from the perspective of $V$ and (b) are true in the state $\psi$.

If $V'\subset V$, then there is a function
\begin{eqnarray}
			\ps\TO^\psi(i_{V'V}):\ps\TO^\psi_V &\longrightarrow& \ps\TO^\psi_{V'}\\
			S_\hP &\longmapsto& S_{\dastoo{V'}{P}}.
\end{eqnarray}
In this way, $\ps\TO^\psi$ becomes a presheaf. It is called the \emph{truth object associated to $\psi$}.

A global element $\ps S$ of $\ps\TO^\psi$ consists of one clopen subset $\ps S_V\in\mc{C}l(\Sig_V)$ for each $V\in\V{}$ such that $\ps S_V|_{V'}=\ps S_{V'}$. Such a global element $\ps S$ clearly is a clopen subobject of $\Sig$. The physical interpretation is that those subobjects $\ps S$ that are global elements of $\ps\TO^\psi$ (and those which are larger than a global element of $\ps\TO^\psi$) represent propositions that are totally true in the state $\psi$. The collection $\Gamma\ps\TO^\psi$ of global elements of the truth object forms a partially ordered set. It is easy to see that this poset is contained in the filter
\begin{equation}
			\Gamma\ps\TO^\psi=\{\ps S\in\Subcl{\Sig} \mid \ps S\geq\ps{\delta(\hP_\psi)}\}.
\end{equation}
The clopen subobject $\ps{\delta(\hP_\psi)}$ hence plays a special r\^ole, it is the smallest subobject representing a totally true proposition. If a classical system is in a pure state $s\in\cS$, then the smallest subset representing a proposition that is true in the state $s$ is $\{s\}$. Hence, the subobject $\ps{\delta(\hP_\psi)}$ is the analogue of a one-element subset $\{s\}$ of the state space $\cS$ of a classical system. 
\begin{equation}
			\wpsi:=\ps{\delta(\hP_\psi)}
\end{equation}
is called the \emph{pseudo-state associated to $\psi$}.

\subsection{Truth-value assignments}
As mentioned before, in classical physics the assignment of truth-values to propositions is straightforward. Given a state $s\in\cS$ of the system, a proposition ``$\Ain\De$'' is \textit{true} if $s$ is contained in the Borel subset $S$ of the state space $\cS$ that represents the proposition, and \textit{false} otherwise. 

In the topos scheme, we have a completely analogous situation: let $\ps S$ be the clopen subobject of the spectral presheaf representing a proposition constructed by conjunction, disjunction and/or negation of elementary propositions ``$\Ain\De$'', and let $\wpsi$ be the pseudo-state associated to some given state $\psi$. It is straightforward to prove that for each $V\in\V{}$,
\begin{equation}
			v(\wpsi\subseteq\ps S)_V=\{V'\subseteq V \mid \wpsi_{V'}\subseteq\ps S_{V'}\}
\end{equation}
is a sieve on $V$. Moreover, if $V'\subset V$, then
\begin{equation}
			v(\wpsi\subseteq\ps S)_{V'}=v(\wpsi\subseteq\ps S)_V\cap\downarrow\!\!V',
\end{equation}
so $v(\wpsi\subseteq\ps S)=(v(\wpsi\subseteq\ps S)_V)_{V\in\V{}}$ is a global element of the subobject classifier $\Om$ of $\SetBHop$, i.e., a topos-internal truth-value for the proposition ``$\Ain\De$'' in the state $\psi$. For more details, see \cite{DI08}. The truth-value $v(\wpsi\subseteq\ps S)$ can be interpreted as the answer of the question `to which degree does the pseudo-state $\wpsi$ lie in the subobject $\ps S$?'. Different from the classical case, where a point $s$ either lies in a subset or not (which determines a Boolean truth-value in the topos $\Set$), we have a truth-value in the logic given by our topos $\SetBHop$.

The simplest description of the truth-value $v(\wpsi\subseteq\ps S)$ is a more global one: $v(\wpsi\subseteq\ps S)$ is the collection of all $V\in\V{}$ such that the component $\wpsi_V$ of the pseudo-state is contained in the component $\ps S_V$ of the subobject representing a proposition. By construction, if $V$ is contained in this collection, then all $V'\subset V$ are also contained in it. A context $V$ is contained in the collection if and only if the local proposition at $V$ is true in the state $\psi$, which is the case if and only if the expectation value of the projection $\hP_{\ps S_V}=\alpha^{-1}(\ps S_V)$ in the state $\psi$ is $1$. %The local proposition at $V$, represented by $\ps S_V$, is obtained from the global proposition that we started from by coarse-graining.

It also becomes clear that the smallest subobjects that can represent totally true propositions are those of the form $\wpsi=\ps{\delta(\hP_\psi)}$. There are smaller non-trivial subobjects $\ps S$, for example those given by a conjunction $\ps{\delta(\hP_{\psi_1})}\wedge\ps{\delta(\hP_{\psi_2})}$. The subobject that represents the proposition ``spin in $x$ direction is up and spin in $y$ direction is down'' is of this form. There is no state $\psi$ that makes this proposition totally true. In this sense, the topos form of quantum logic takes care of the conjunction of non-compatible propositions in a non-trivial way, different from Birkhoff-von Neumann quantum logic. We conjecture that this feature will lead to a logical formulation of the uncertainty relations.

Each pure state $\psi$ determines a truth-value assignment
\begin{eqnarray}
			v_\psi:\Subcl{\Sig} &\longrightarrow& \Gamma\Om\\			\nonumber
			\ps S &\longmapsto& v(\wpsi\subseteq \ps S).
\end{eqnarray}
Since both the clopen subobjects $\Subcl{\Sig}$ of the spectral presheaf and the truth-values $\Gamma\Om$ form a Heyting algebra, one might wonder if $v_\psi$ is a homomorphism of Heyting algebras. Let $\ps S_1,\ps S_2$ be two clopen subobjects. Then, for all $V\in\V{}$,
\begin{eqnarray*}
			v_\psi(\ps S_1\wedge\ps S_2)_V &=& \{V'\subseteq V \mid \hP_{(\ps S_1\wedge\ps S_2)_V}\geq\hP_\psi\}\\
			&=& \{V'\subseteq V \mid \hP_{\ps S_{1;V}}\wedge\hP_{\ps S_{2;V}}\geq\hP_\psi\}\\
			&=& \{V'\subseteq V \mid \hP_{\ps S_{1;V}}\geq\hP_\psi\}\cap\{V'\subseteq V \mid \hP_{\ps S_{2;V}}\geq\hP_\psi\}\\
			&=& v_\psi(\ps S_1)_V\wedge v_\psi(\ps S_2)_V,
\end{eqnarray*}
so we obtain
\begin{equation}
			\forall \ps S_1,\ps S_2\in\Subcl{\Sig}:v_{\psi}(\ps S_1\wedge\ps S_2)=v_\psi(\ps S_1)\wedge v_\psi(\ps S_2).
\end{equation}
Truth-value assignments thus preserve conjunction. On the other hand, for all $V\in\V{}$,
\begin{eqnarray*}
			v_\psi(\ps S_1\vee\ps S_2)_V &=& \{V'\subseteq V \mid \hP_{(\ps S_1\vee\ps S_2)_V}\geq\hP_\psi\}\\
			&=& \{V'\subseteq V \mid \hP_{\ps S_{1;V}}\vee\hP_{\ps S_{2;V}}\geq\hP_\psi\}\\
			&\supseteq& \{V'\subseteq V \mid \hP_{\ps S_{1;V}}\geq\hP_\psi\}\cup\{V'\subseteq V \mid \hP_{\ps S_{2;V}}\geq\hP_\psi\}\\
			&=& v_\psi(\ps S_1)_V\vee v_\psi(\ps S_2)_V,
\end{eqnarray*}
so
\begin{equation}
			\forall \ps S_1,\ps S_2\in\Subcl{\Sig}:v_{\psi}(\ps S_1\vee\ps S_2) \geq v_\psi(\ps S_1)\vee v_\psi(\ps S_2).
\end{equation}
A truth-value assignment $v_\psi$ need not preserve disjunction. In general, the truth-value of a disjunction of two propositions is larger than the disjunction of the truth-values of the propositions. Clearly, this relates to superposition. While we have used a formulation of the truth-value assignment employing projections, one could as well formulate everything just using clopen subsets of Gel'fand spectra. This shows that the topos form of quantum logic preserves that part of standard quantum logic that relates to superposition, but without the need for linear structures.

%%%%%%%%%%%%%%%%%%%%%%%%%%%%%%%%%%%%%%%%%%%%%%%%%%%%%%%%%%%%%%%%%%%%%%%%%%%%%%%%%%%%%%%%%%%%%%%%%%%%%%%%%%%%%%%%%%
\section{Mixed states}
In this short section, we will sketch how mixed states can be treated in the topos approach and how they relate to the logical aspects.

%%%%%%%%%%%%%%%%%%%%%%%%%%%%%%%%%%%%%%%%%%%%%%%%%%%%%%%%%%%%%%%%%%%%%%%%%%%%%%%%%%%%%%%%%%%%%%%%%%%%%%%%%%%%%%%%%%
\subsection{States as measures on the spectral presheaf}
Let $\rho$ be an arbitrary state of the quantum system under consideration. $\rho$ is a positive linear functional on $\BH$ of norm $1$. This is very different from the classical case, where an arbitrary state is a \emph{probability measure} $\mu$ on the state space $\cS$ of the system.

Interestingly, the topos approach allows the representation of arbitrary states $\rho$ of a quantum system by probability measures on the spectral presheaf, as was shown in \cite{Doe08}. We refer to this article for the proofs of the results in this section.

The \emph{measure $\mu_\rho$ associated to $\rho$} is the mapping
\begin{eqnarray}			\label{Defmu_rho}
			\mu_\rho:\Subcl{\Sig} &\longrightarrow& \Gamma\ps{[0,1]^\succeq}\\	\nonumber	
			\ps S=(\ps S_V)_{V\in\V{}} &\longmapsto& \mu_\rho(\ps S)=(\rho(\hP_{\ps S_V}))_{V\in\V{}}.
\end{eqnarray}
The codomain $\Gamma\ps{[0,1]^\succeq}$ denotes antitone functions from $\V{}$ to the unit interval $[0,1]$, i.e., if $g\in\Gamma\ps{[0,1]^\succeq}$ and $V'\subset V$, then $1\geq g(V')\geq g(V)\geq 0$. (These functions can be understood as the global elements of a certain presheaf, hence the notation.) To each clopen subobject, such a function is assigned by the measure $\mu_\rho$. It is straightforward to see that $\mu_\rho(\Sig)=1_{\V{}}$, the function that is constantly $1$.

The abstract definition of a measure is as follows: a mapping
\begin{eqnarray}			\label{Defmu}
			\mu:\Subcl{\Sig} &\longrightarrow& \Gamma\ps{[0,1]^\succeq}\\
			\ps S=(\ps S_V)_{V\in\V{}} &\longmapsto& \mu(\ps S)=(\mu(\ps S_V))_{V\in\V{}}
\end{eqnarray}
is called a \emph{measure on the clopen subobjects of $\Sig$} if the following two conditions are fulfilled:
\begin{itemize}
	\item $\mu(\Sig)=1_{\V{}}$;
	\item for all $\ps S_1,\ps S_2\in\Subcl{\Sig}$, it holds that $\mu(\ps S_1\vee\ps S_2)+\mu(\ps S_1\wedge\ps S_2) =\mu(\ps S_1)+\mu(\ps S_2)$.
\end{itemize}
Somewhat surprisingly, these very weak conditions---which do not refer to non-commutativity or linearity in any direct sense---suffice to determine a unique state $\rho_\mu$ on $\BH$ provided $\rm{dim}(\Hi)\geq 3$, and every state arises that way. Measures on the spectral presheaf hence completely encode positive linear functionals on the algebra $\BH$ of physical quantities.

%%%%%%%%%%%%%%%%%%%%%%%%%%%%%%%%%%%%%%%%%%%%%%%%%%%%%%%%%%%%%%%%%%%%%%%%%%%%%%%%%%%%%%%%%%%%%%%%%%%%%%%%%%%%%%%%%%
\subsection{The relation between measures and logical aspects}
Let $\rho=\psi$ be a pure state, and let $\mu_\psi$ be the corresponding measure. Clearly, if for some $\ps S\in\Subcl{\Sig}$ we have $\mu_\psi(\ps S)=1_{\V{}}$, i.e., the clopen subobject $\ps S$ is of measure $1_{\V{}}$, then $\ps S$ represents a proposition that is totally true in the state described by $\mu_\psi$. The smallest subobject of measure $1_{\V{}}$ with respect to the measure $\mu_\psi$, that is, the support of the measure $\mu_\psi$, is the pseudo-state $\wpsi$.

More generally, the truth-value of the proposition represented by $\ps S$ in the state represented by $\mu_\psi$ is the collection of all those $V\in\V{}$ such that $\mu_\psi(\ps S)(V)=1$, since
\begin{eqnarray*}
			v_\psi(\ps S) &=& \{V\in\V{} \mid \mu_\psi(\ps S)(V)=1\}\\
			&=& \{V\in\V{} \mid \bra\psi\hP_{\ps S_V}\ket\psi=1\}\\
			&=& \{V\in\V{} \mid \hP_{\ps S_V}\geq\hP_\psi\},
\end{eqnarray*}
where $\hP_{\ps S_V}=\alpha^{-1}(\ps S_V)$. The measure $\mu_\psi$ corresponding to a pure state $\psi$ hence encodes the logical aspects given by the truth-value assignment $v_\psi$ determined by $\psi$.

For mixed states, there is no such simple connection between the measure-theoretical and the logical aspects. The pseudo-state $\wpsi$ corresponding to a pure state $\psi$ determines a unique measure $\mu_\psi$ that has $\wpsi$ as its support. In contrast to that, a mixed state $\rho$ is not determined uniquely by its support. One may, however, describe an arbitrary mixed state $\rho$ uniquely in terms of a family of generalised truth objects.

Instead of considering a single truth object $\ps\TO^\rho$, we define a family $(\ps\TO^\rho_r)_{r\in(0,1]}$ by
\begin{eqnarray}
			\forall V\in\V{} \forall r\in(0,1]:\ps\TO^\rho_{r;V} &:=& \{S\in\mc{C}l(\Sig_V) \mid \rho(\hP_S)\geq r \}\\
			&=& \{S\in\mc{C}l(\Sig_V) \mid \mu_\rho(S)\geq r \}.
\end{eqnarray}
This is a direct generalisation of the definition of a truth object (equation \eq{DefTO}), since
\begin{eqnarray*}
			\ps\TO^\psi_V &=& \{S_\hP\in\mc{C}l(\Sig_V) \mid \hP\geq\hP_\psi\}\\
			&=& \{S\in\mc{C}l(\Sig_V) \mid \hP_S\geq\hP_\psi\}\\
			&=& \{S\in\mc{C}l(\Sig_V) \mid \bra\psi\hP_S\ket\psi=1\}\\
			&=& \{S\in\mc{C}l(\Sig_V) \mid \bra\psi\hP_S\ket\psi\geq 1\}		
\end{eqnarray*}
so the truth object $\ps\TO^\psi$ is equal to the element $\ps\TO^\psi_1$ of the family $(\ps\TO^\psi_r)_{r\in(0,1]}$.

It is easy to see that each $\ps\TO^\rho_r,\;r\in(0,1],$ is a presheaf over $\V{}$. The global elements of $\ps\TO^\rho_r$ are clopen subobjects that are of measure $r_{\V{}}$ or greater with respect to the measure $\mu_\rho$. Here, $r_{\V{}}$ is the function that is constantly $r$ on $\V{}$.

We saw that every measure $\mu_\rho$ determines a unique family $(\ps\TO^\rho_r)_{r\in(0,1]}$ of generalised truth objects. Conversely, the measure $\mu_\rho$ can be reconstructed from the family $(\ps\TO^\rho_r)_{r\in(0,1]}$. This and many further aspects will be treated in detail in \cite{DI09}.

%%%%%%%%%%%%%%%%%%%%%%%%%%%%%%%%%%%%%%%%%%%%%%%%%%%%%%%%%%%%%%%%%%%%%%%%%%%%%%%%%%%%%%%%%%%%%%%%%%%%%%%%%%%%%%%%%%
\section{Related work}
Recently, Landsman et al. proposed to another scheme of topos quantum logic that is closely related to ours \cite{HLS09,CHLS09}. The main difference lies in the fact that Landsman et al. are using the topos $\SetBH$ of \emph{covariant} functors over the context category. In fact, in \cite{HLS09}, they consider an arbitrary $C^*$-algebra, not just $\BH$. Yet, in \cite{CHLS09}, where some definitions from \cite{HLS09} are made explicit, the special case of the algebra $M_n(\bbC)$ is used, which of course equals $\BH$ for an $n$-dimensional Hilbert space. The main advantage of using covariant functors is that the external algebra $\BH$ determines a canonical internal algebra $\fu\BH$ which is an \emph{abelian} $C^*$-algebra in the topos $\SetBH$. By constructive Gel'fand duality, as developed by Banaschewski and Mulvey \cite{BanMul97,BanMul00,BanMul00b,BanMul06}, this internal algebra has a Gel'fand spectrum $\Sgm$, which is a \emph{locale} in the topos $\SetBH$. (A locale is a generalised topological space, see e.g. \cite{Jst02}.) Landsman et al. suggest to use the opens in this locale as the representatives of propositions. For a detailed comparison between the contravariant and the covariant approach, see \cite{Doe10b}.

The use of a form of intuitionistic logic for quantum theory has also been suggested by Coecke in \cite{Coe02}. He uses a construction discovered by Bruns and Lakser, the so-called injective hull of meet-semilattices (see also \cite{Stu05}) to embed a meet-semilattice of propositions into a Heyting algebra by introducing new joins to the meet-semilattice. There are no further obvious connections between this approach and the topos form of quantum logic, but it would be interesting to compare both constructions with respect to the underlying geometric structures: both approaches formulate an intuitionistic form of quantum logic using Heyting algebras, and every complete Heyting algebra is a locale and hence a generalised topological space.

%%%%%%%%%%%%%%%%%%%%%%%%%%%%%%%%%%%%%%%%%%%%%%%%%%%%%%%%%%%%%%%%%%%%%%%%%%%%%%%%%%%%%%%%%%%%%%%%%%%%%%%%%%%%%%%%%%
\section{Conclusion}
We presented the main features of the topos form of quantum logic. This new form of quantum logic is distributive, intuitionistic, multi-valued and contextual. Moreover, it has a clear underlying geometric structure. The spectral presheaf serves as an analogue of the state space of a classical system, with propositions being represented as subobjects. Interestingly, topos quantum logic preserves that part of standard quantum logic that relates to superposition. The truth-value assignments given by pure states do not depend on any notion of measurement and observers, i.e., there is no need for an instrumentalist interpretation. Mixed states can be fully described as measures on the spectral presheaf. There are interesting relations between the measure-theoretical and the logical aspects of the theory that will be further investigated in future work.

\textbf{Acknowledgements.} First of all, I want to thank Chris Isham for many discussions, constant support and his great generosity. I thank Bob Coecke, Pedro Resende, Chris Mulvey, Steve Vickers and Bas Spitters for useful discussions. I also want to thank the organisers of QPL'08 and '09 for creating a very interesting workshop.

\end{document}